\documentclass[11pt]{iopart}
\usepackage{graphicx,cite}
\usepackage{epsf,amsfonts,amssymb,amsbsy}
\begin{document}
\title[Gravitational lensing\ldots by vector field] {Gravitational lensing due
to dark matter modelled by vector field}
\author{V.V.Kiselev*$\raisebox{1mm}{\dag}$\,
and D.I.Yudin$\raisebox{1mm}{\dag}$}
\address{*
\ Russian State Research Center ``Institute for High Energy
Physics'', 
Pobeda 1, Protvino, Moscow Region, 142281, Russia}
\address{\dag\ 
Moscow Institute of Physics and Technology, Institutskii per. 9,
Dolgoprudnyi Moscow Region, 141700, Russia}
\ead{kiselev@th1.ihep.su}
\begin{abstract}
The specified constant 4-vector field reproducing the spherically
symmetric stationary metric of cold dark matter halo in the region
of flat rotation curves results in a constant angle of light
deflection at small impact distances. The effective deflecting
mass is factor $\pi/2$ greater than the dark matter mass. The
perturbation of deflection picture due to the halo edge is
evaluated.
\end{abstract}

\section{Introduction}

For spherically symmetric stationary dark halos in spiral galaxies
one could explore the weak field approximation of pressureless
matter implying the metric
\begin{equation}\label{dark_metric}
    {\rm d}s^2=\big\{1+2\,\phi_\textsc{\small gr}(r)\big\}{\rm d}t^2-
    \frac{{\rm d}r^2}{\displaystyle 1-\frac{2 G M_\textsc{\small dm}(r)}{r}}-r^2[{\rm
    d}\theta^2+\sin^2\theta{\rm d}\varphi^2],
\end{equation}
where $M_\textsc{\small dm}(r)$ denotes the dark matter mass
inside the sphere of radius $r$, since to linear order in small
$M_\textsc{\small dm}(r)/r$ and $\phi_\textsc{\small gr}(r)$ the
Einstein equations give
\begin{equation}\label{Ein1}
    8\pi G\rho(r)=G_t^t=R_t^t-\frac{1}{2}\,g_t^t\,R=\frac{2G M'_\textsc{\small
    dm}(r)}{r^2}
\end{equation}
exactly yielding
\begin{equation}\label{mass}
    \int\limits_0^r{\rm d}^3\boldsymbol r\,\rho(r)=M_\textsc{\small
    dm}(r),
\end{equation}
while $\phi_\textsc{\small gr}(r)$ represents the gravitational
potential of dark mass
\begin{equation}\label{poten}
    \phi_\textsc{\small gr}(r)=\int\limits^r{\rm
    d}r\,\frac{G M_\textsc{\small dm}(r)}{r^2}
\end{equation}
producing the force
\begin{equation}\label{force}
    \boldsymbol F=\boldsymbol \nabla\phi_\textsc{\small gr}(r)=-
    \frac{G M_\textsc{\small dm}(r)}{r^2}\,\frac{\boldsymbol r}{r}
\end{equation}
as well as a negligible pressure
\begin{equation}\label{pressure}
    8\pi G\,p_n(r)=-G_n^n\approx 0,\qquad n=\{r,\theta,\varphi\}.
\end{equation}
For flat rotational curves with the characteristic velocity
$v_0^2$ we get
\begin{equation}\label{flat}
    \frac{v_0^2}{r}=\frac{G M_\textsc{\small dm}(r)}{r^2}
\end{equation}
and, hence,
\begin{equation}\label{M_v}
    M_\textsc{\small dm}(r)=\frac{1}{G}\,v_0^2 r,
\end{equation}
which argues for the metric
\begin{equation}\label{static}
    {\rm d}s^2 =\texttt{f}(r)\,{\rm d}t^2-\frac{1}
    {\texttt{h}(r)}\,{\rm d}r^2-r^2[{\rm
    d}\theta^2+\sin^2\theta\,{\rm d}\varphi^2],
\end{equation}
with
\begin{equation}
    \texttt{h}(r) =
    1-2v_0^2,\quad\texttt{f}'(r)=\frac{2v_0^2}{r}.
\end{equation}
We consider the metric is close to the Minkowskian one up to small
corrections in $v_0^2\to 0$, so that
\begin{equation}\label{imtegral_f}
    \texttt{f}(r)=1+2v_0^2\ln r. 
\end{equation}

Surprisingly, metric (\ref{static}) was derived for the dark
matter modelled by 4-vector field $\mathcal{A}_\nu$ with constant
components \cite{K}
\begin{equation}\label{4-vector}
    \mathcal{A}_0=\frac{1}{\sqrt{8\pi G}},\qquad
    \boldsymbol {\mathcal{A}}=\frac{v_0^2}{\sqrt{4\pi G}}\,
    \frac{\boldsymbol r}{r}.
\end{equation}
In \cite{vghost} we supposed a treating the vector field in terms
of ghost condensate
\cite{Arkani-Hamed,Mukohyama,Mukohyama2,Krause,Piazza,Peloso,Dubovsky,Krotov}:
 $\partial_t\phi=\dot \phi=\mathcal{A}_0$, and
global spherically symmetric monopole $\boldsymbol
\psi(\boldsymbol r)=\boldsymbol
\nabla\chi(r)=\boldsymbol{\mathcal{A}}(\boldsymbol r)\sim
\boldsymbol n=\boldsymbol r/r$, combined into the 4-vector
$\mathcal{A}_\nu$. The phenomenon of ghost condensate is an
analogue of
    Higgs mechanism for the condensation of scalar field with negative
    square of mass near zero values of field. This mechanism of ghost condensation
    usefully works in the case of expanding homogeneous universe, that
    has a specific reference frame with a selected temporal component in comparison
    with spatial ones. Then, the scalar field with negative sign
    of kinetic term near zero values of kinetic energy in its action
    composed of invariant derivative terms $X=\partial_\mu\phi\partial^\mu\phi$,
    condenses
    at a nonzero value of $\dot\phi$, which represents a temporal component of
    4-vector field. The ghost condensation removes the problem with negative
    kinetic energy, since near the condensate the fluctuations
    have correct sign of kinetic energy, of course. The studies \cite{K}
    show, that field $\dot\phi$ should be constant due to the current evolution of universe,
    with a high accuracy. Next, one could combine the ghost condensate with the
    global monopole, that produces a correct static spherically symmetric
    profile of mass distribution
    $\rho\sim 1/r^2$ in the region of flat rotation curves due to the dark
    matter. Therefore, one should suppose the presence of global
    monopole in the center of spiral galaxy, i.e. inside a
    supermassive black hole.
    At large distances, the monopole corresponds to the spatial component of vector field.
    Then, one considers the background field $\phi=\mu_\ast^2 t-\mu_\star^2 r$
    at large distances (perturbations become important at small distances,
    corresponding to inner edge of halo), where $\mu_\ast$ is the scale
    about the Planck mass, while the ratio of $\mu_\star/\mu_\ast$ determines
    the rotation velocity $v_0$ (see details in \cite{vghost}).
    So, the ghost condensate in the presence of global monopole with account of
    higher derivative terms in the action can be represented by the constant 4-vector
    field $\mathcal{A}_\mu$ relevant to the situation of both universe expansion and
    dark halo.

In the present paper we calculate the light bending by the metric
modelling the dark halo. In Section 2 we in detail calculate the
deflection angle in the limit of `infinite' halo implying a
negligible value of impact distance with respect to the halo size
and compare the result with the deflection by a global monopole
\cite{Vilenkin}. Then we determine the effective deflecting mass
and find its difference by a factor of $\pi/2$ from the dark
matter mass responsible for the flat rotation curves in contrast
to naive expectations of their coincidence. Further, we describe
the geometry of deflection revealing properties consistent with
astronomical observations. In Section 3 we calculate the
deflection in the case of finite halo, that demonstrates a weak
dependence on the impact distance slowly perturbing the geometry
discussed in Section 2. The results are summarized in Conclusion.

\section{Limit of infinite halo}

\begin{figure}[th]
\centerline{  \includegraphics[width=8cm]{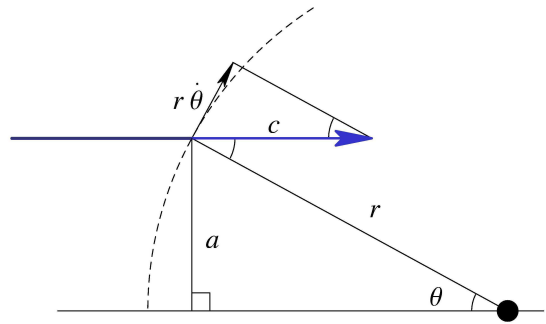}}

  \caption{Geometrical relations between the impact distance $a$ and
  circular velocity $r\dot \theta$.}\label{fall}
\end{figure}

By the Hamilton--Jacobi equation
\begin{equation}\label{1}
    g^{\mu\nu}\;{\partial_{\mu} S}\,{\partial_{\nu} S}  =
    0
\end{equation}
for an action $S$ of massless particle written in the form
\begin{equation}\label{2}
    S = -{\cal E}\, t+{\mathfrak M}\,\theta+{\cal S}(r),
\end{equation}
incorporating two integrals of motion in the spherically symmetric
static gravitational field at fixed polar angle
$\phi=\mbox{const.}$, so that $\cal E$ is the conserved energy of
particle and $\mathfrak M$ is its rotational momentum, which we
put positive for definiteness, we deduce
\begin{equation}\label{3}
    \left(\frac{\partial {\cal S}}{\partial r}\right)^2 =
    \frac{1}{\texttt{f}\texttt{h}}\,{\cal E}^2-\frac{1}{\texttt{h}}
    \,\frac{\mathfrak
    M^2}{r^2},
\end{equation}
which results in
\begin{equation}\label{4}
    {\cal S}(r) = \pm\int \limits_{r_0}^{r(t)} \textrm{d}r\;
    \frac{1}{\sqrt{\texttt{f}\texttt{h}}}
    \sqrt{{\cal E}^2-V^2(r)},
\end{equation}
where $V^2$ is an analogue of centripetal potential,
\begin{equation*}
V^2(r) = \texttt{f}\,\frac{\mathfrak M^2}{r^2}.
\end{equation*}
Thus, the quantity ${\cal S}(r)$ represents the action
    for the radial motion of particle in the gravitational field
    with account of energy due to the angular rotation.
The trajectory is implicitly determined by the
equations\footnote{Here we take the positive sign of square root
for brevity of record.}
\begin{eqnarray}
  \frac{\partial S}{\partial {\cal E}} &=& \textsf{const} = -t
  +\int\limits_{r_0}^{r(t)} \textrm{d}r\; \frac{1}{\sqrt{\texttt{f}\texttt{h}}}
    \frac{\cal E}{\sqrt{{\cal E}^2-V^2(r)}}, \label{p1}\\
  \frac{\partial S}{\partial {\mathfrak M}} &=& \textsf{const} =
  \theta
  -\int\limits_{r_0}^{r(t)} \textrm{d}r\;\frac{1}{\sqrt{\texttt{f}\texttt{h}}}\,
  \frac{\texttt{f}}{r^2}
    \frac{\mathfrak M}{\sqrt{{\cal E}^2-V^2(r)}}.\label{p2}
\end{eqnarray}
Taking the derivative of (\ref{p1}) and (\ref{p2}) with respect to
the time\footnote{As usual $\partial_t f(t) =\dot f$.}, we get
\begin{eqnarray}
  1 &=& \dot r\; \frac{\cal E}{\sqrt{\texttt{f}\texttt{h}}\,
  \sqrt{{\cal E}^2-V^2(r)}}, \label{9}\\
  \dot \theta &=& \frac{\dot
  r}{r^2}\;\frac{\texttt{f}}{\sqrt{\texttt{f}\texttt{h}}}\,
  \frac{\mathfrak M}{  \sqrt{{\cal
  E}^2-V^2(r)}},\label{10}
\end{eqnarray}
and, hence,
\begin{equation}\label{8}
    {\cal E} = \frac{\texttt{f}}{r^2 \dot \theta}\,{\mathfrak M},
\end{equation}
relating the energy and rotational momentum. Let us denote the
ratio of two integrals by
\begin{equation*}
    \frac{\mathfrak M}{\mathcal E}=a
\end{equation*}
with the dimension of length, which we call as `an impact
distance', that becomes clear from Fig. \ref{fall} shown for the
flat Minkowskian space-time.

At the trajectory given by (\ref{9}) and (\ref{10}) we get
\begin{equation}\label{rd}
    \frac{{\rm d}\theta}{{\rm
    d}r}=\mp\frac{a}{r^2}\,\sqrt{\frac{\texttt{f}}{\texttt{h}}}\,\frac{1}
    {\displaystyle\sqrt{1-\frac{a^2\,\texttt{f}}{r^2}}},
\end{equation}
where the negative sign corresponds to the branch of trajectory
for the particle arriving from infinity, while the positive sign
does to the branch for the particle departing to infinity. Denote
the solution of
\begin{equation*}
    r^2=a^2\texttt{f}(r)
\end{equation*}
by $r_{\rm min}$. At this value the radial velocity $\dot r$
becomes equal to zero (the denominator in Eq.(\ref{rd}) vanishes),
that corresponds to returning point. So, the light ray approaches
the gravitating object from the infinity, while it reaches the
distance $r_{\rm min}$, after that the ray moves off the object.
Therefore, the quantity $r_{\rm min}$ denotes the minimal distance
between the light trajectory and the gravitational lens. Then, the
overall change of $\theta$ for the particle is given by
\begin{equation}\label{delta-t}
    \Delta \theta_G= 2\int\limits_{r_{\rm min}}^{\infty}\,{\rm d} r
    \,\frac{a}{r^2}\,\sqrt{\frac{\texttt{f}}{\texttt{h}}}\,\frac{1}
    {\displaystyle\sqrt{1-\frac{a^2\,\texttt{f}}{r^2}}},
\end{equation}
if we put the size of halo $r_\star$ much greater than the impact
distance: $a^2\texttt{f}/r_\star^2\to 0$, that corresponds to the
limit of infinitely large halo. In contrast to \cite{NSS}, we
expect that this limit is relevant to actual astronomical
observations.

 Let us
integrate out (\ref{delta-t}) by expanding in small parameter of
$v_0^2\to 0$. So, with the accuracy up to the first order we get
\begin{equation*}
    \texttt{f}= \texttt{f}(r_{\rm
    min})+2v_0^2\ln\frac{r}{r_{\rm min}}+\mathcal{O}(v_0^4),
\end{equation*}
\begin{equation*}
    \texttt{h}= 1-2v_0^2,
\end{equation*}
that allows us to write down
\begin{equation}\label{expand}
    \Delta \theta_G=\Delta\theta_0+v_0^2(\Delta\theta_1+\Delta\theta_2+
    \Delta\theta_3)+\mathcal{O}(v_0^4),
\end{equation}
where the leading term
\begin{equation}\label{t0}
    \Delta\theta_0=2\int\limits_{r_{\rm min}}^{\infty}\,{\rm d} r
    \,\frac{a}{r^2}\,\sqrt{\texttt{f}(r_{\rm min})}\,\frac{1}
    {\displaystyle\sqrt{1-\frac{a^2\,\texttt{f}(r_{\rm min})}{r^2}}}
\end{equation}
after the substitution
\begin{equation}\label{z}
    z=\frac{a}{r}\,\sqrt{\texttt{f}(r_{\rm min})}
\end{equation}
is reduced to
\begin{equation}\label{t0-1}
    \Delta\theta_0=2\int\limits_0^1\frac{{\rm
    d}z}{\sqrt{1-z^2}}=2\arcsin z\Big|_0^1=\pi,
\end{equation}
which corresponds to the straightforward motion with no
deflection. The contribution $\Delta \theta_1$ originates from the
factor of $1/\sqrt{\texttt{h}}\approx 1+v_0^2$ in the integrand of
(\ref{delta-t}), so that
\begin{equation}\label{t1}
    \Delta\theta_1=\Delta\theta_0=\pi.
\end{equation}
The expansion of
\begin{equation*}
    \sqrt{\texttt{f}}= \sqrt{\texttt{f}(r_{\rm
    min})}-v_0^2\ln z+\mathcal{O}(v_0^4)
\end{equation*}
results in
\begin{equation}\label{t2-1}
    \Delta\theta_2=-2\int\limits_0^1{\rm
    d}z\,\frac{\ln z}{\sqrt{1-z^2}},
\end{equation}
where we have neglected the deviation of $\texttt{f}(r_{\rm min})$
from unit
\begin{equation*}
    \texttt{f}(r_{\rm min})=1+\mathcal{O}(v_0^2).
\end{equation*}
Analogously, the expansion of
\begin{equation*}
    \frac{1}
    {\displaystyle\sqrt{1-\frac{a^2\,\texttt{f}}{r^2}}}=
    \frac{1}{\sqrt{1-z^2}}-v_0^2\,\frac{z^2\ln
    z}{\sqrt{(1-z^2)^3}}+\mathcal{O}(v_0^4)
\end{equation*}
produces
\begin{equation}\label{t3}
    \Delta\theta_3=-2\int\limits_0^1{\rm
    d}z\,\frac{z^2\ln z}{\sqrt{(1-z^2)^3}}.
\end{equation}
Summing up (\ref{t2-1}) and (\ref{t3}) gives
\begin{equation}\label{t23}
    \Delta\theta_2+\Delta\theta_3=-2\int\limits_0^1{\rm
    d}z\,\frac{\ln
    z}{\sqrt{(1-z^2)^3}}.
\end{equation}
The integral is easily calculated by parts
\begin{equation*}
    -2\int\limits{\rm
    d}z\,\frac{\ln z}{\sqrt{(1-z^2)^3}}=
        -\frac{2z}{\sqrt{1-z^2}}\,\ln z+2
    \int\limits\frac{{\rm d}z}{\sqrt{1-z^2}},
\end{equation*}
so that
\begin{equation}\label{t23-1}
    \Delta\theta_2+\Delta\theta_3=\pi.
\end{equation}
Thus, the deflection of light is determined by the angle
\begin{equation}\label{deflect}
    \Delta\theta=\Delta\theta_G-\pi=
    v_0^2(\Delta\theta_1+\Delta\theta_2+\Delta\theta_3)=2\pi v_0^2.
\end{equation}
Note, this limit of deflection angle is independent of impact
distance. This result is similar to the calculation of light
deflection by the global monopole metric in \cite{Vilenkin}, where
the radial component $\texttt{h}$ can be taken in the same form,
while the temporal component of metric is constant,
$\texttt{f}\mapsto\mbox{const.}$, which corresponds to the term of
$\Delta\theta_1$ in calculations above. Therefore, we emphasize
that \textit{the deflection angle in the metric of global monopole
is twice less than the calculated value of} $\Delta\theta=2\pi
v_0^2$.

The bending angle derived is in agreement with results of
\cite{Kar}, where the deflection due to the dark matter with a
significant pressure was calculated numerically at various
equation of state parameters $w_n=p_n/\rho$, and the limit of
pressureless matter was mentioned, too. The same result could be
extracted from \cite{NSS} by calculating the limit of $v_0^2\to 0$
at infinitely small impact distances from analytical expressions
in \cite{NSS} also discussed in Section 3.

\subsection{Deflecting dark matter}

The main feature of deflection angle under study is its
independence of impact distance. Such the deflection can be
ascribed to an effective dark matter mass. Indeed, let us compare
the effect with the deflection by a nonrelativistic spherical mass
distribution described by the Schwarzschild metric:
\begin{equation}\label{schwarz}
    \Delta\theta_{BH}=\frac{4M G}{r_{\rm min}}.
\end{equation}
Putting
\begin{equation}\label{eff}
    \Delta\theta=\frac{4\widetilde M_\textsc{dm}(r_{\rm min}) G}{r_{\rm min}},
\end{equation}
we derive the expression for the effective dark matter mass
responsible for the deflection
\begin{equation}\label{eff_M}
    \widetilde M_\textsc{dm}(r)=\frac{\pi}{2G}\,v_0^2\,r,
\end{equation}
which \textit{linearly depends on the distance}. Compare it with
the dark matter mass responsible for the centripetal acceleration
in the region of flat rotational curves,
\begin{equation}\label{rot}
    \frac{v_0^2}{r}=\frac{GM_\textsc{dm}(r)}{r^2}\quad
    \Rightarrow\quad
    M_\textsc{dm}(r)=\frac{1}{G}\,v_0^2\,r.
\end{equation}
Therefore, in the metric under consideration the effective
deflecting mass of dark matter is factor $\pi/2$ greater than the
dark matter mass producing flat rotational curves in Newtonian
limit.

The same conclusion is drawn from virial estimates of dark matter
mass in galaxy clusters, where we get the relation
\begin{equation*}
    \langle v^2\rangle =\frac{G\langle
    M_\textsc{dm}\rangle}{2\langle r\rangle},
\end{equation*}
since the average kinetic energy of $N$ galaxies with average
masses $\langle M_{\rm galaxy}\rangle =\langle
M_\textsc{dm}\rangle/N$ at $N\gg 1$
\begin{equation*}
    \langle T_{\rm kin}\rangle =\frac{1}{2}\,N\,\langle
    M_{\rm galaxy}\rangle\langle v^2\rangle=\frac{1}{2}\,\langle
    M_\textsc{dm}\rangle\langle v^2\rangle
\end{equation*}
is related with the average potential energy of galaxies posed at
average distance $\langle r\rangle$
\begin{equation*}
    \langle V\rangle =-\frac{N^2}{2}\,\frac{G\langle M_{\rm
    galaxy}\rangle ^2}{\langle
    r\rangle}=-\frac{1}{2}\,\frac{G\langle
    M_\textsc{dm}\rangle^2}{\langle r\rangle}
\end{equation*}
by the virial theorem
\begin{equation*}
    2\langle T_{\rm kin}\rangle+\langle V\rangle=0.
\end{equation*}
Both estimates produce the similar linear dependence of dark
matter mass versus the distance, as we retrieve from the
gravitational lensing.

Remarkably, astronomical observations of galaxy clusters give
\begin{equation}\label{lumy}
    \langle M_\textsc{dm}\rangle \sim 200\langle
    M_\textsc{b}\rangle,
\end{equation}
where $\langle M_\textsc{b}\rangle$ is the mass of luminous
baryonic matter in the clusters. Note, that the mass of luminous
matter is related with the velocity of rotation in spiral galaxies
at distances of flat rotational curves by the empirical
Tully--Fisher scaling law
\begin{equation}\label{TF}
    \langle M_\textsc{b}\rangle \sim v_0^4,
\end{equation}
which is natural, if one suggests the existence of universal
critical Milgrom acceleration \cite{Milgrom}
\begin{equation}\label{Milgrom}
    a_0=\frac{v_0^2}{r_0}=\frac{G \langle
    M_\textsc{b}\rangle}{r_0^2},
\end{equation}
where $r_0$ is a characteristic distance for a galaxy, where the
dominance of baryonic matter in the gravitation comes to
competition with the dark matter contribution becoming dominant at
larger distances. The value of $r_0$ is close to the visible size
of galaxy, i.e. it is the characteristic size of baryon
distribution in the galaxy. So, let us compare the masses of
luminous and dark matters:
\begin{equation}\label{comare}
    \langle M_\textsc{b}\rangle \simeq \frac{1}{G}\,v_0^2\,r_0,
    \qquad
    \langle M_\textsc{dm}\rangle \simeq
    \frac{1}{G}\,v_0^2\,r_\star,
\end{equation}
where $r_\star$ is the size of dark halo. Therefore, we conclude
that
\begin{equation}\label{com_rr}
    r_\star\sim 200\,r_0,
\end{equation}
i.e. the dark halo spreads far away from the visible size of
luminous object. This point supports the relevance of considering
the limit of infinite halo in the problem of gravitational lensing
at realistic impact distances about the visible size of lens.


Thus, the vector field model of dark matter qualitatively
reproduces both dark matter effects for astronomical objects: the
flat rotational curves and light bending. The numerical effect on
about 50\%-difference between the effective deflecting mass and
dark matter gravitational mass in rotational curves could probably
be used in future for falsifying the model, though such a
combination of astronomical data seems to be quite exclusive.

Another aspect of comparing the rotation curve picture in a galaxy
with the gravitational lensing by the same galaxy has been
recently considered in \cite{Visser}, where the potentials
responsible for the rotation and lensing have been introduced. The
difference between the potentials would signalize on a non-zeroth
pressure of dark matter. In the case under study these potentials
coincide, so the lensing of dark mass distribution could be
evaluated by standard techniques for the cold pressureless matter.
In contrast to the potentials, we operate by the effective
deflecting mass allowing us to compare it with the mass one could
expect from the rotation velocity. This could be important for a
treatment of some misalignment between the lensing mass and dark
matter mass in astronomical measurements of X-band lenses
\cite{obser-X}.

\subsection{Geometry of deflection}

Let us consider the geometric aspects of gravitational lensing. In
Fig. \ref{bending}, point $S$ denotes a star, which light reaches
an observer posed in point $O$ by path $SBO$ curved by dark matter
object somewhere placed at interval $BD$. Since all of angles are
extremely small, while distances between the observer, dark matter
object and star are astronomically large, we can perform
calculations in the first order versus angles. Then, the distance
between the deflection point and position of dark matter object
projected to the line between the observer and star $r=BD$ (see
Fig. \ref{bending}) equals
\begin{equation}\label{r}
    r=l\,\alpha;\qquad r=d\,\beta.
\end{equation}
Points $D$ and $D'$ coincide because $|DD'|\approx
l\,\alpha^2=\mathcal{O}(\alpha^2)$ is negligible. The angle of
deflection $\Delta \theta$ is independent of $r$ and
\begin{equation}\label{sum}
    \Delta \theta=\alpha+\beta.
\end{equation}
Therefore,
\begin{equation}\label{delta}
    \beta=\Delta\theta\;\frac{l}{l+d}.
\end{equation}
The same deflection takes place by the path in the upper half of
plane shown in Fig. \ref{bending}.
\setlength{\unitlength}{1mm}
\begin{figure}[th]
    \begin{center}
    \begin{picture}(120,60)
    \put(0,0){\includegraphics[width=12cm]{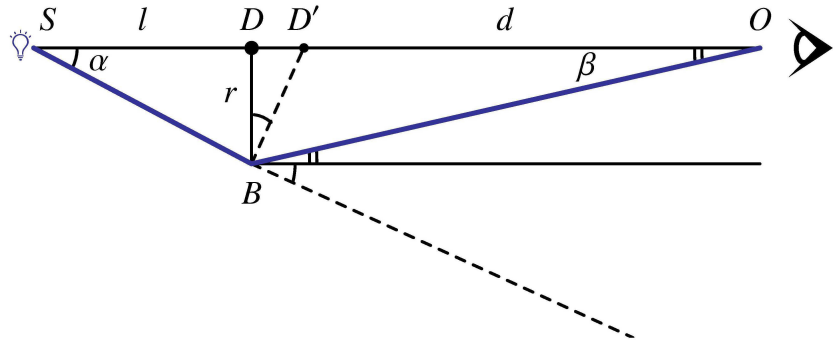}}
    \end{picture}
    \end{center}
  \caption{The plane geometry of light bending.}\label{bending}
\end{figure}

\begin{figure}[th]
  \centerline{\includegraphics[width=12cm]{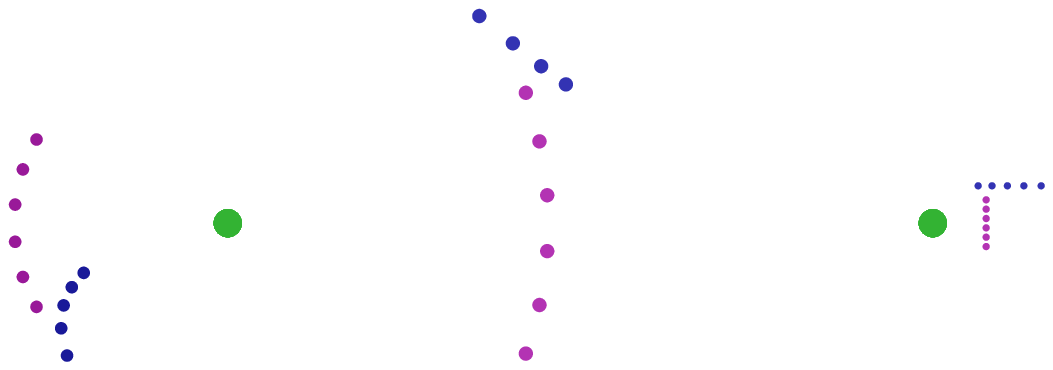}}
  \caption{The gravitational lensing of linear image by the dark mater object
  (\textit{left}) in comparison with the initial pattern without the light bending
  (\textit{right}).}\label{bend}
\end{figure}
\begin{figure}[th]
  \centerline{\includegraphics[width=9cm]{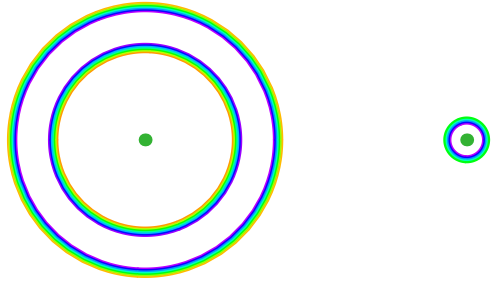}}
  \caption{The gravitational lensing of centered circular image by the dark mater object
  (\textit{left}) in comparison with the initial pattern without the light bending
  (\textit{right}).}\label{bend6}
\end{figure}
\noindent Thus, the angle between two images of star is given by
\begin{equation}\label{delta2}
    \Delta \beta=2\beta=2\,\Delta\theta\;\frac{l}{l+d}.
\end{equation}
This geometry can be easily applied to draw samples of
gravitational lensing in the model with constant $\Delta \theta$.

So, Fig. \ref{bend} represents the gravitational magnifying of
linear image at $d=1$, $l=10$ and small angles. The picture
illustrates the characteristic distortion caused by such lenses.
For simplicity of picturing, in Figs. 3-6 we put
    the parameter $v_0^2=1/(40\pi) \approx 0.008\approx (0.09)^2$ in order to enlarge
    the visual effect. Nevertheless, in practice, an angular magnification of
    real astronomical facilities is sufficiently large for the observation of
    gravitational lensing by the dark matter (see numerical
    estimates in the end of section).

One can easily recognize basic features of deflection and
distortion:
\begin{enumerate}
    \item The dark matter object is posed in the center of
    inversion symmetry.
    \item Hence, the angle distances between the points inside two
    images coincide. (If one displaces the star image by angle
    $\gamma$, then the second image is displayed to the same angle
    increment $\gamma$.)
    \item Therefore, small radial widths of images are coincide,
    while the radial sizes are proportional. The coefficient of
    proportionality $\kappa$ is equal to the ratio of angle fractions given by
    division of $\Delta \beta$ by the center of symmetry, i.e. the
    dark matter object. This point is shown in Fig. \ref{bend6}.
    \item Then, the ratio of surface areas is equal to $\kappa$ as
    illustrated in Fig. \ref{bend2} for the distortion of circle object.
\begin{figure}[th]
  \centerline{\includegraphics[width=12cm]{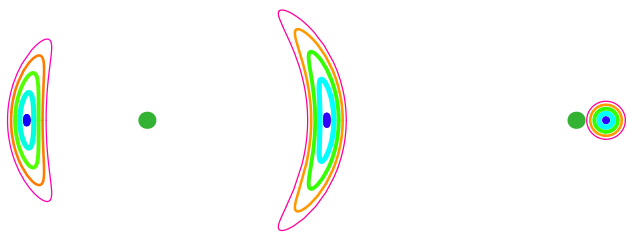}}
  \caption{The gravitational lensing of circular image by the dark mater object
  (\textit{left}) in comparison with the initial pattern without the light bending
  (\textit{right}).}\label{bend2}
\end{figure}
\end{enumerate}

Thus, for the model of gravitational lensing, we arrive to the
following conclusion:

The relative values of surface brightness for two images of lensed
small object is given the inverse ratio of angle fractions,
$\kappa$, or visible impact parameters.

\begin{figure}[th]
  \centerline{\includegraphics[width=12cm]{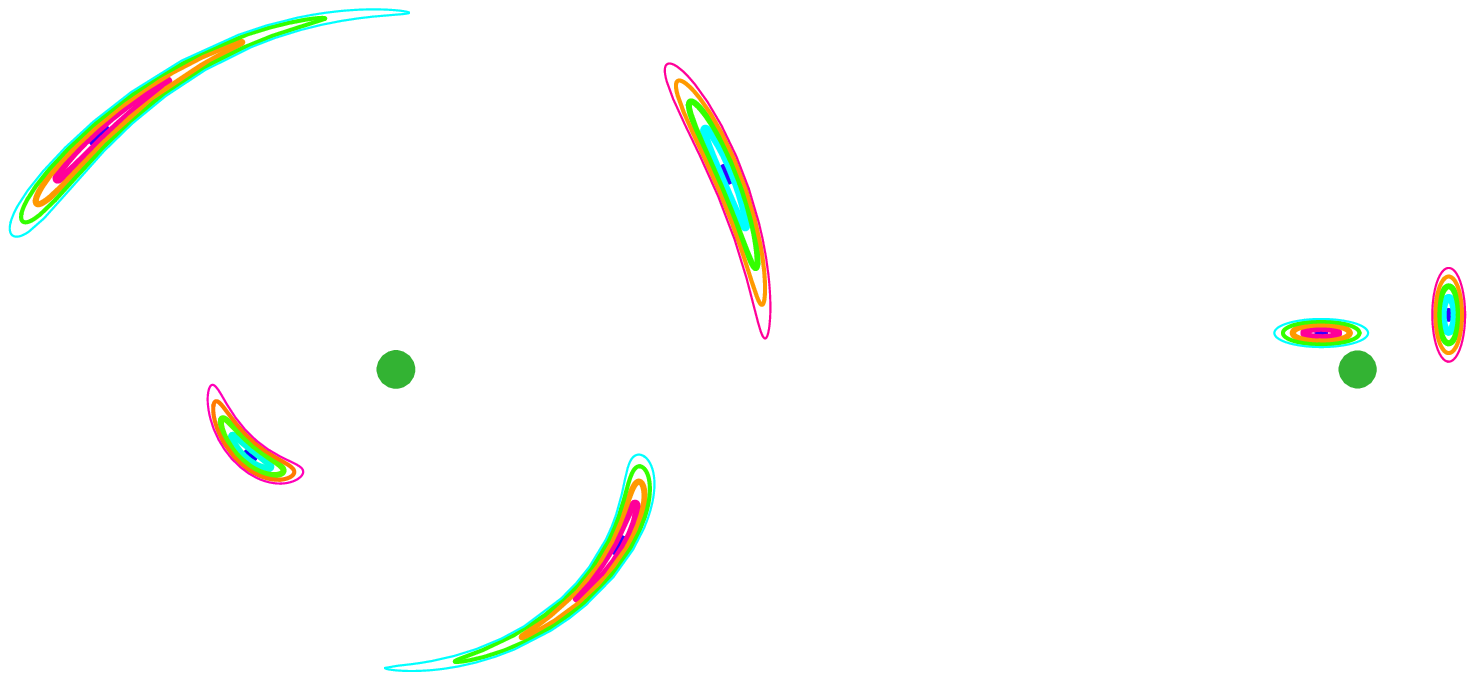}}
  \centerline{\hspace*{5mm}\includegraphics[width=12cm]{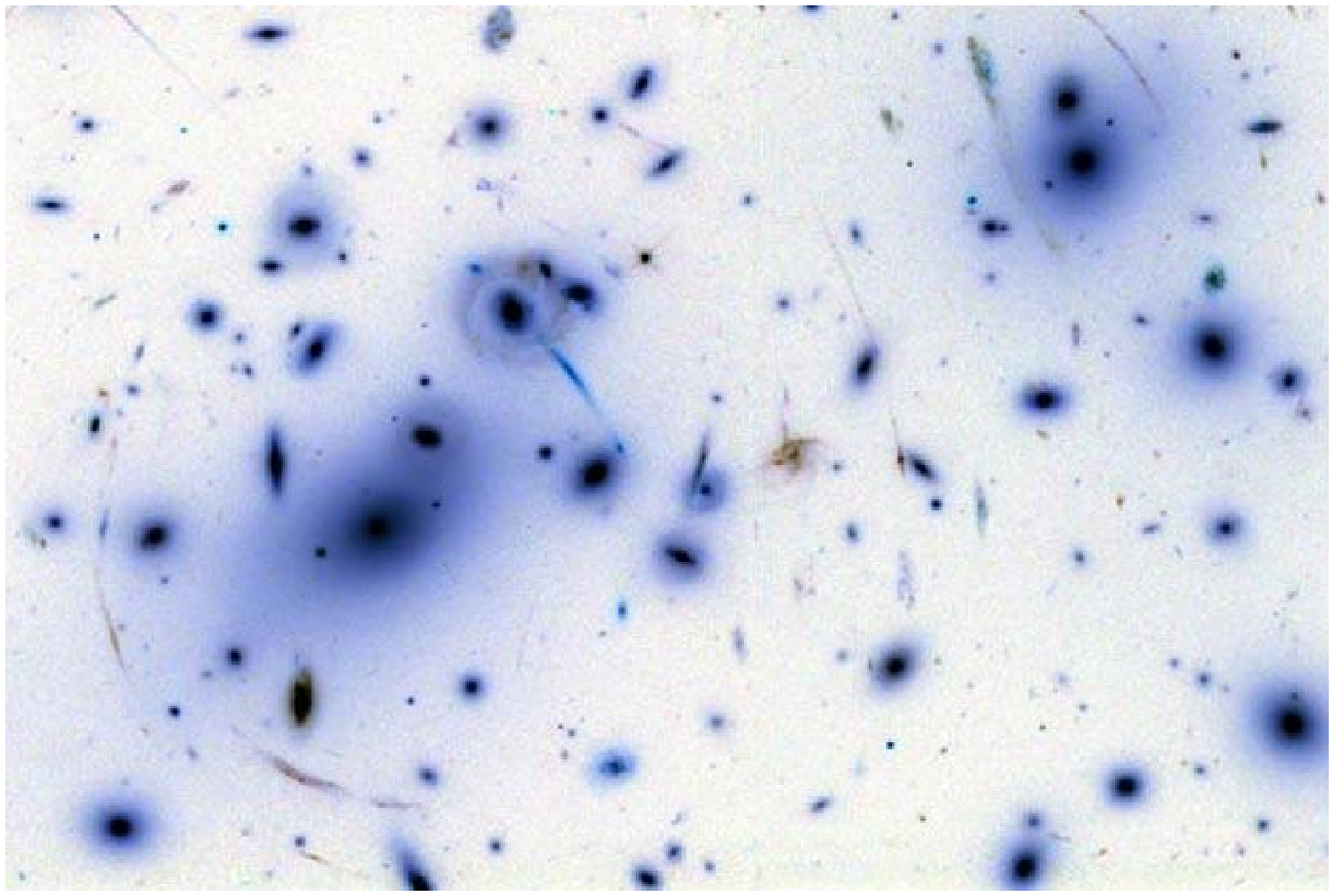}}
  \caption{The gravitational lensing of sample galaxies at different distances
  by the dark mater object (\textit{top-left}) in comparison with
  the initial pattern without the light bending
  (\textit{top-right}) and with the real picture of Abell 2218 cluster lensing
  (\textit{bottom}) [from NASA, 1998].}\label{bend5}
\end{figure}

So, the ordinary picture of gravitational lensing due to the dark
matter modelled by the vector field is shown in Fig. \ref{bend5}.
We have taken two galaxies as elliptic figures at different
distances ($d=1$, $l_1=9$, $l_2=3$). Their images are narrow arcs
surrounding the dark matter object, that is generically similar to
the real astronomical picture inserted as illustration.

\subsection{Numerical values}
The characteristic angle between two images for an object due to
the gravitational lensing is about
\begin{equation}\label{num1}
    \Delta \beta \simeq 2\Delta \theta=4\pi v_0^2\approx 2.6\cdot
    10^6\,v_0^2\,\mbox{arcsec.}
\end{equation}
A typical velocity of rotation is in the range of $100-400$
km/sec, i.e. $v_0^2\sim 10^{-6}$. Then,
\begin{equation}\label{num2}
    \Delta\beta\sim 2\,\mbox{arcsec.}
\end{equation}
Such the value is consistent with astronomical observations of
gravitational lensing, where the characteristic angle sizes
between the dark matter object and image of lensed luminous object
is about 1 arcsec.

\section{Edge of halo}

\begin{figure}[th]
\centerline{  \includegraphics[width=11cm]{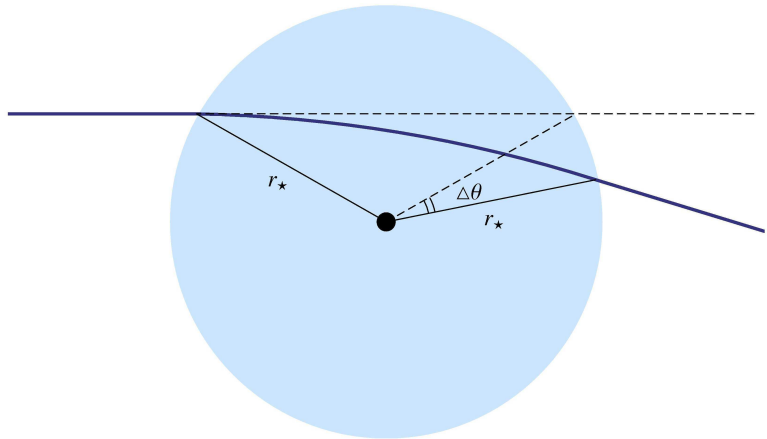}}

  \caption{The propagation of light through the dark halo of finite size $r_\star$.
  }\label{edge}
\end{figure}

If the impact distance is comparable with the size of dark halo,
then one has to take into account the halo edge at $r=r_\star$
(see Fig. \ref{edge}). So, the formulae exploited above give the
angle increment in the halo,
\begin{equation}\label{D_halo}
    \Delta\theta_G(r_\star)=\pi-2\arcsin z_\star+v_0^2 \left(
    2\pi-4\arcsin z_\star+2\,\frac{z_\star\ln
    z_\star}{\sqrt{1-z_\star^2}}
    \right),
\end{equation}
with $z_\star={a}/{r_\star}$.

Let us suppose that beyond the halo we get the Schwarzschild
metric with the gravitational radius $r_g$:
$\texttt{f},\texttt{h}\mapsto 1-\frac{r_g}{r}$. Then, the
increment of angle beyond the halo is determined by the integral
\begin{equation}\label{delta-Schwarz}
    \Delta \theta_\textsc{bh}(r_\star)= 2\int\limits_{r_\star}^{\infty}\,{\rm d} r
    \,\frac{a}{r^2}\,\frac{1}
    {\displaystyle\sqrt{1-\frac{a^2}{r^2}\left(1-\frac{r_g}{r}\right)}},
\end{equation}
which can be easily evaluated after the substitution
\begin{equation*}
    r=\rho-\frac{1}{2}\,r_g,
\end{equation*}
and expansion to the first order in $r_g/a$. So, we get
\begin{equation}\label{D-Sc}
    \Delta\theta_\textsc{bh}(r_\star)\approx
    2\int\limits_{r_\star}^\infty\frac{{\rm d}\rho}{\rho^2}\,
    \frac{a\displaystyle\left(1+\frac{r_g}{\rho}\right)}{\displaystyle
    \sqrt{1-\frac{a^2}{\rho^2}}}\,
    =
    2\arcsin
    z_\star+2\,\frac{r_g}{a}\left(1-\sqrt{1-z_\star^2}\right).
\end{equation}
The value of gravitational radius could be fixed by the dark
matter mass at $r_\star$,
\begin{equation}\label{rg}
    r_g=2GM_\textsc{dm}(r_\star)=2v_0^2r_\star.
\end{equation}
Then, the overall deflection angle
\begin{equation*}
    \Delta
    \theta_\star=\Delta\theta_G(r_\star)+\Delta\theta_\textsc{bh}(r_\star)-\pi
\end{equation*}
is given by
\begin{equation}\label{d_star}
    \Delta\theta_\star=v_0^2\left\{2\pi-4\arcsin z_\star+2\,\frac{z_\star\ln
    z_\star}{\sqrt{1-z_\star^2}}+\frac{4}{z_\star}\left(1-\sqrt{1-z_\star^2}\right)
    \right\}
\end{equation}
valid at $0\leqslant z_\star\leqslant 1$, while at $z_\star> 1$ we
get the ordinary angle of bending
$\Delta\theta_\textsc{bh}=2r_g/a$.

\begin{figure}[th]
    \begin{center}
  \begin{picture}(145,60)
  \put(-10,0){  \includegraphics[width=8cm]{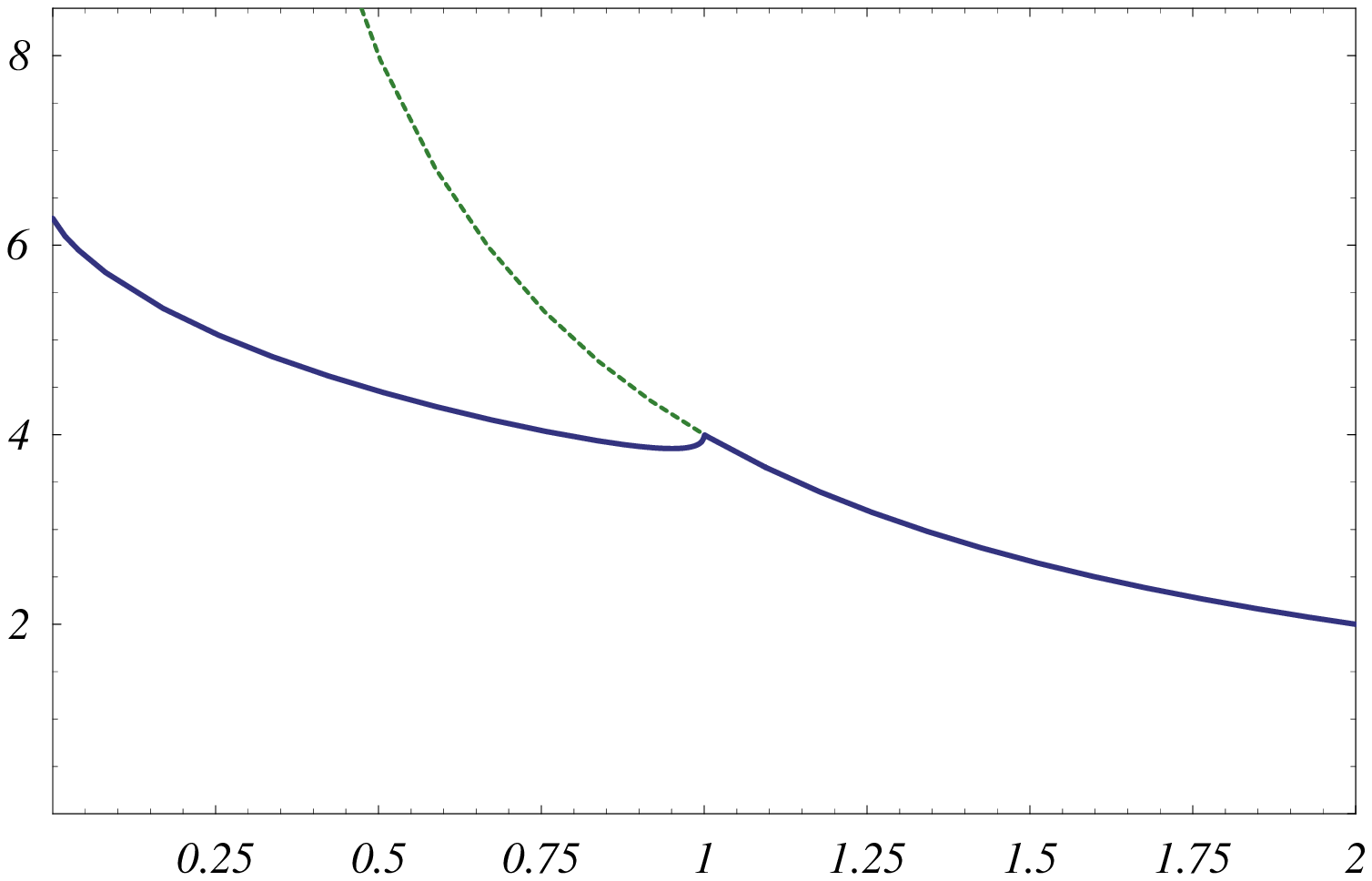}}
  \put(75,0){  \includegraphics[width=8cm]{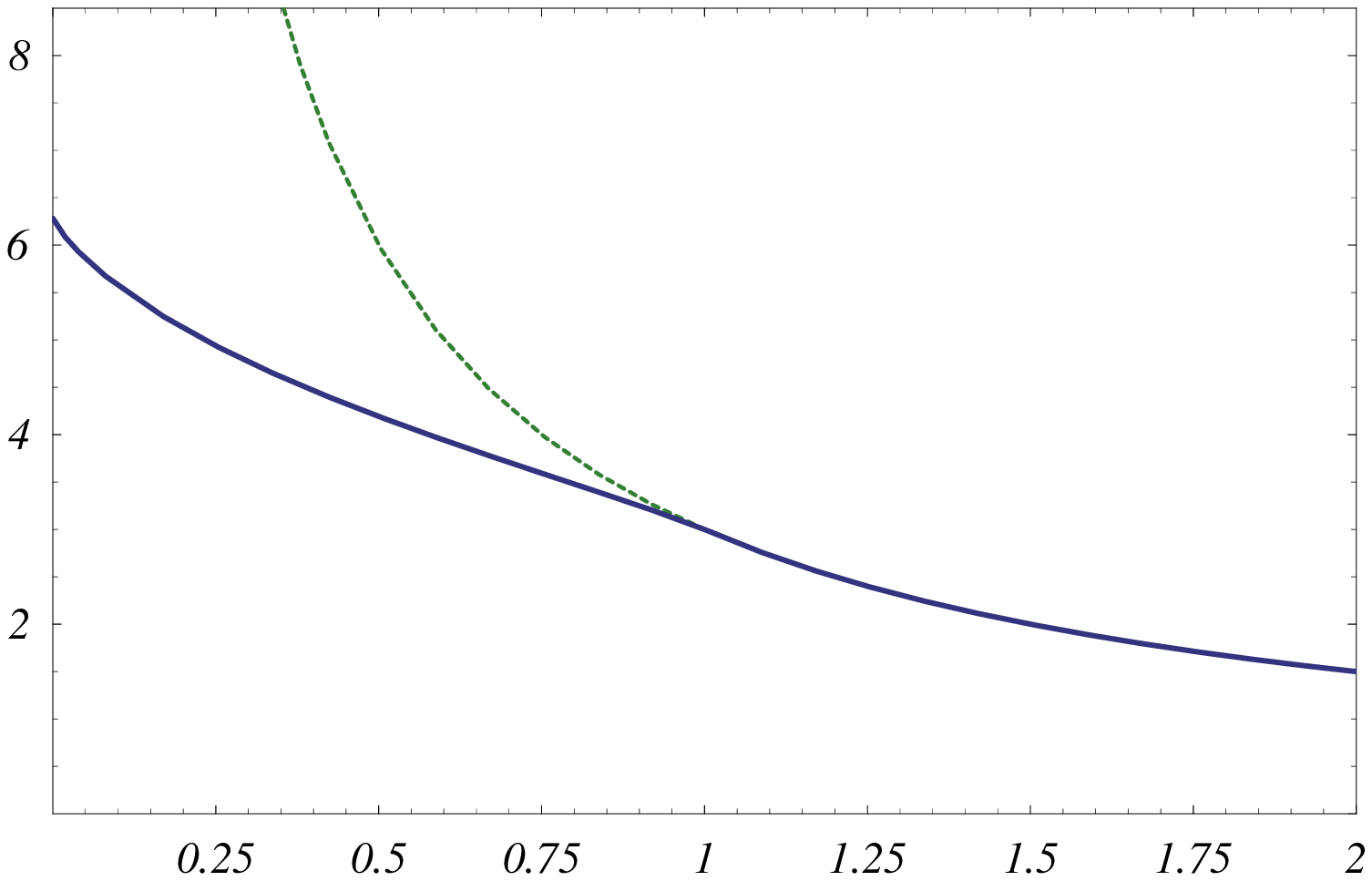}}
  \put(65,-3){$z_\star$}
  \put(150,-3){$z_\star$}
  \put(-5,53){$\Delta\theta_\star/v_0^2$}
  \put(80,53){$\Delta\theta_\star/v_0^2$}
  \end{picture}
  \end{center}
  \caption{The deflection angle light by the dark matter with the edge at
  $r=r_\star$ versus $z_\star=a/r_\star$
  under matching with the Schwarzschild metric at $r_g=2v_0^2r_\star$ (\textit{left}) and
  $r_g=3v_0^2r_\star/2$ (\textit{right}) in comparison with the deflection by the massive point
  of $M=r_g/(2G)$
  (\textit{dotted line}).}\label{match}
\end{figure}

Note, that the derivative of deflection angle has a singularity at
$z_\star\to 1$,
\begin{equation}\label{dt/dz}
    \frac{{\rm d}\Delta\theta_\star}{{\rm
    d}z_\star}\to\frac{v_0^2}{\sqrt{1-z_\star^2}}\,\left(\frac{2r_g}{v_0^2r_\star}-3
    \right),\qquad z_\star\to 1.
\end{equation}
The singularity reveals as a fracture point of curve in Fig.
\ref{match}. The fracture disappears at
$r_g=\frac{3}{2}v_0^2r_\star$. However, such the redefinition of
parameter $r_g$, removing the fracture of bending curve (the
derivative of $\Delta\theta_\star$ with respect to $z_\star$
becomes a continuous curve), results in a discontinuity of metric
coefficient $\texttt{h}$ at the edge of halo, that is absolutely
unacceptable.

The origin of singularity in Eq.(\ref{dt/dz}) is the specific
dependence of halo metric-coefficient $\texttt{f}$ on the distance
$r$. So, it produces a slow deflection with negative sign in
vicinity of halo edge ($z_\star=1$), i.e. there is a small
repulsive component of force acting to the light ray in comparison
with the case of point-like mass. It is clearly seen in derivation
of Eqs. (\ref{t3}) and (\ref{t23}). Then, the attraction by the
halo near the edge is small in comparison with the case of
point-like mass, hence, the substitution of trajectory passing the
halo instead of the same path in the field of point-like mass
results in an essentially smaller deflection, and we even get the
fracture of curve for the bending angle, while the deflection
increases when the path in the halo becomes longer. Thus, the
fracture is the direct consequence of metric, specified for the
dark halo. Moreover, one could speak about a gravitational
refraction of light on the border of two media with different
metrics (in the case under study, we get the transition from the
vacuum to the halo medium).

The analysis shows that in quite a wide region of impact factors,
the deflection by the dark halo with the edge is slowly perturbed
due to the edge: the dependence on the impact distance is weak in
comparison with the rapid change in the case of point-like mass.
So, the picture of deflection considered in Section 2
qualitatively remains the same.

\paragraph{Comparison with the paper by Nucamendi, Salgado,
Sudarsky} \hspace*{-2mm}\cite{NSS}. In \cite{NSS} authors in part
considered the same problem of light bending by the dark halo with
the edge. The difference is twofold. First, the metric coefficient
$\texttt{f}$ in \cite{NSS} has been taken in the form
\begin{equation}\label{NSS1}
    \texttt{f}=A\left(\frac{r}{r_0}\right)^{2v_0^2}.
\end{equation}
Second, the deflection by the Schwarzschild metric calculated by
expanding in powers of inverse distance has been written in a
form, which is literally different, but in fact it should
coincides with the formula (\ref{D-Sc}) with the accuracy up to
$\mathcal{O}(r_g^2/a^2)$. These facts allow us to make a direct
comparison in the limit of $v_0^2\to 0$. So, we find that our
result is consistent with the corresponding limit of expression in
\cite{NSS} with the accuracy mentioned above. In part, the limit
of infinite halo $z_\star\to 0$ in \cite {NSS}, reproduces the
value of $\Delta \theta=2\pi v_0^2$, which is actually in the
evident contradiction with the main result of \cite{NSS} shown
graphically as the function of $z_\star$ (see Fig. 1 in
\cite{NSS}). Indeed, instead of $2\pi$ one could observe a number,
which is factor $4/3$ greater. The same factor remains, when we
compare the complete curve in \cite{NSS}: the deflection angle of
dark halo with the edge is multiplied by $4/3$. Literally, the
curve of \cite{NSS} could be reproduced from our \textit{right}
curve in Fig. \ref{match} multiplied by $4/3$, that gives the
correct value for the bending by the Schwarzschild metric, but it
misleads on the deflection by the halo. The factor $4/3$ removes
the fracture point from the curve, of course, as it is clearly
seen in Fig. 1 in \cite{NSS}.

Thus, our analysis on the deflection angle by the dark halo with
edge is consistent with the formal limit of expressions in
\cite{NSS}, but we disagree with the numerical in-figure-shown
result of \cite{NSS}.

\paragraph{An inner edge.} We have considered the external edge of
halo at large distances. However, the luminous matter essentially
contributes to the deflection at short impact distances, where the
gravitational field is dominantly determined by distributions of
stars and gas. So, there is the inner edge of halo. This problem
has been comprehensively investigated in terms of Modified
Newtonian Dynamics (MOND) \cite{Milgrom}. Recently, Bekenstein
formulated a consistent relativistic theory of gravitation
including gravitational scalar and vector fields in addition to
the ordinary tensor of metric \cite{Bekenstein}, so that it
incorporates general MOND effects originated from the introduction
of critical acceleration $a_0$. The authors of \cite{MOND_lensing}
have found the same value of deflection for the case of infinite
halo at impact parameters greater than the inner edge, that
corresponds to the isothermal halo of circular velocity $v_0$. The
comparison with observational data has been performed, so that the
conclusion drawn is the following: MOND is not in conflict with
the data on the gravitational lensing. Then, the analysis of
\cite{MOND_lensing} supports the model considered in this paper,
which includes general features of MOND, too.

\section{Conclusion}

Briefly summarizing the consideration, we have described the
deflection of light by the dark mater halo modelled by the
constant 4-vector field, in detail. The analytical results are in
agreement with the limit of small velocity of rotation $v_0^2\to
0$ in previous studies.

The deflection angle for the infinite halo is twice of the value
for the global monopole. The effective lensing mass is factor
$\pi/2$ greater than the dark matter mass at the same distance.
For the typical rotation velocity in spiral galaxies, the
characteristic value of deflection angle is about several
arcseconds, that is consistent with values observed
astronomically. The relative surface brightness of two images of
lensed object is determined by the inverse ratio of impact angles
with respect to the dark lens.

The effect of halo edge is reduced to slow dependence of
deflection angle on the impact distance, that weakly perturbs the
gravitational lensing.

In conclusion, we have to make several important comments:

    First, it would be wrong to identify the model of halo accepted in the paper
    with known model of isothermal halo.
    Such the opinion could not be correct because of the following reasons:
    \begin{itemize}
    \item[i)] The physics of vector field model in the paper and the
    physics of general isothermal model are certainly different.
    \item[ii)] The vector field model of dark matter includes
    inner and outer edges, so that between the edges the metric is
    described by the same metric as for the isothermal halo.
    Hence, the deflection at impact distances greater than the
    radius of inner edge and much less that the radius of outer
    edge coincides with the result of isothermal halo. That is a
    new statement for the limit of infinite halo in the vector
    field model with different physics. The final equation for the
    deflection is known \cite{Falco}, but the
    physical motivation is different. For instance, the model does
    not contain the singularity of isothermal halo (that is quite
    evident). So, the identification with the singular
    isothermal halo looks misleading. 
    \item[iii)] The effects related with the inner edge absent in
    the case of isothermal halo were investigated in the
    literature \cite{MOND_lensing}.
    \item[iv)] The effects related with the outer edge
    absent in the case of isothermal halo as was represented in the
    literature \cite{NSS},
    contain some errors, i.e.
    the presence of fracture point in the bending curve at the edge was missed.
    The representation of correction in the paper has required
    further application of method,
    which gives the correct result for the infinite halo, that has been usefully
    made in the initial part of Section 2.
    \end{itemize}
    Thus, the paper describes the physics different from the singular isothermal
    halo. The results concerning the limit of infinite halo in the vector field
    model and the effects of outer edge of halo are new.

    Second, we have to emphasize that the reason for
    the difference between the effective mass deflecting the light
    at the impact distance $r$ and the dark matter mass inside the same radius
    $r$ by a factor found in the paper is quite clear:
    the deflection takes place due to the mass inside
    the radius $r$ and effectively due to the dark matter mass outside the same radius.
    This reason is evident and transparent.

This work is partially supported by the grant of the president of
Russian Federation for scientific schools NSc-1303.2003.2, and the
Russian Foundation for Basic Research, grant 04-02-17530.

\end{document}